\documentclass[12pt]{article}
\usepackage{bibunits}
\defaultbibliographystyle{naturemag}
\defaultbibliography{ref}
\usepackage{times}
\usepackage{fixltx2e}
\usepackage{graphicx}
\usepackage[table]{xcolor}
\usepackage{ragged2e}
\usepackage{multirow}
\usepackage{helvet}
\usepackage{booktabs}
\usepackage{colortbl}
\usepackage{amsmath}
\usepackage{bm}
\usepackage{cmbright}

\usepackage{hyperref}
\hypersetup{
    colorlinks=true,
    linkcolor=magenta,
    filecolor=magenta, 
    citecolor=sared,
    urlcolor=red,
}

\usepackage{blindtext}

\makeatletter
\newenvironment{figurehere}
{\def\@captype{figure}}
{}
\makeatletter

\usepackage{ccaption}
\usepackage{siunitx}
\usepackage{fixltx2e}
\newenvironment{sciabstract}{%
\begin{quote} \bf}
{\end{quote}}

\definecolor{sared}{rgb}{0.69, 0,0}

\newcounter{lastnote}

\title{\textbf{Probabilistic Tree Inference Enabled by FDSOI Ferroelectric FETs
}} 

\author{
Pengyu Ren$^{1}$, Xingtian Wang$^{1}$, Boyang Cheng$^{1}$, Jiahui Duan$^{1}$, Giuk Kim$^{1}$\\
Xuezhong Niu$^{1}$, Halid Mulaosmanovic$^{2}$, Stefan Duenkel$^{2}$, Sven Beyer$^{2}$\\
X. Sharon Hu$^{1}$, Ningyuan Cao$^{1}$, Kai Ni$^{1}$\\
\normalsize{$^{1}$University of Notre Dame, Notre Dame, IN 46556, USA}\\
\normalsize{$^{2}$GlobalFoundries Fab1 LLC \&Co. KG, Dresden 01109, Germany}
}

\date{}

\begin{document} 

\maketitle 

\vspace{5mm}
\begin{sciabstract}

Artificial intelligence applications in autonomous driving, medical diagnostics, and financial systems increasingly demand machine learning models that can provide robust uncertainty quantification, interpretability, and noise resilience. Bayesian decision trees (BDTs) are attractive for these tasks because they combine probabilistic reasoning, interpretable decision-making, and robustness to noise. However, existing hardware implementations of BDTs based on CPUs and GPUs are limited by memory bottlenecks and irregular processing patterns, while multi-platform solutions exploiting analog content-addressable memory (ACAM) and Gaussian random number generators (GRNGs) introduce integration complexity and energy overheads. Here we report a monolithic FDSOI-FeFET hardware platform that natively supports both ACAM and GRNG functionalities. The ferroelectric polarization of FeFETs enables compact, energy-efficient multi-bit storage for ACAM, and band-to-band tunneling in the gate-to-drain overlap region and subsequent hole storage in the floating body provides a high-quality entropy source for GRNG. System-level evaluations demonstrate that the proposed architecture provides robust uncertainty estimation, interpretability, and noise tolerance with high energy efficiency. Under both dataset noise and device variations, it achieves over 40\% higher classification accuracy on MNIST compared to conventional decision trees. Moreover, it delivers more than two orders of magnitude speedup over CPU and GPU baselines and over four orders of magnitude improvement in energy efficiency, making it a scalable solution for deploying BDTs in resource-constrained and safety-critical environments.

\end{sciabstract}

\section*{\textcolor{sared}{\Large Introduction}}

% [OUTLINE: 1. Senarios need 3 properties uncertainty estimation, interpretability, noise robustness. 2. BDT, the combination of DT and BNN is good for it 3. Among different implementations of BDT, CPU and GPU is no efficient enough, need ACAM and GRNG for it 4. There are different platforms of ACAM and GRNG, and RRAM， FEFET, FDSOI-FEFET single platform for BDT, FDSOI-FEFET is the best 5.]

% \section{Introduction}
Recent advancements in machine learning have profoundly transformed multiple critical application domains, including autonomous driving, medical diagnosis, pharmaceutical drug development, and financial investment. In autonomous driving, robust uncertainty estimation, explainability, and noise resilience are crucial for ensuring safety-critical decision-making under complex and unpredictable conditions \cite{grigorescu2020,feng2021}. Similarly, in medical diagnosis and drug development, accurate uncertainty quantification significantly enhances diagnostic reliability, assisting clinicians in making informed and correct decisions \cite{esteva2019,rajpurkar2022}. In the financial sector, these capabilities are vital for risk assessment and investment decisions, enabling algorithms to adapt effectively to market volatility and reducing financial risks \cite{dixon2020,fischer2018}.

Several machine learning models have been proposed to address these challenges, broadly including basic neural network-based models and tree-based models. Neural network models are capable of learning complex nonlinear representations from large-scale data and have demonstrated strong performance across various tasks \cite{bonnet2023bringing,esteva2017dermatologist}. In contrast, tree-based models are widely valued for their inherent interpretability and efficient decision-making mechanisms, making them attractive for applications where model explainability is required  \cite{lundberg2020local, pedretti2021}, particularly in risk-critical applications. Traditional decision trees provide clear and interpretable decision structures, but typically lack effective uncertainty estimation and robustness to noise, limiting their applicability in scenarios where prediction confidence is critical \cite{nakahara2025}. On the other hand, Bayesian neural networks (BNNs) incorporate probabilistic inference and can naturally quantify prediction uncertainty while maintaining strong noise tolerance \cite{lin2023uncertainty}. However, their multilayered architectures often make them difficult to interpret. To combine the advantages of these two paradigms, bayesian decision tree(BDT) has emerged as a promising hybrid approach. BDT integrates the interpretable hierarchical structure of decision trees with Bayesian inference, enabling probabilistic predictions and improved robustness to uncertainty and noise by modeling each tree node threshold as a Gaussian distribution and sampling from it during each inference \cite{nakahara2025,nuti2021}. As a result, BDT provides a balanced model that simultaneously offers interpretability and reliable uncertainty estimation (Fig.~\ref{fig:motivation}a).

\begin{figurehere}
   \centering
    \includegraphics[scale=1,width=1\textwidth]{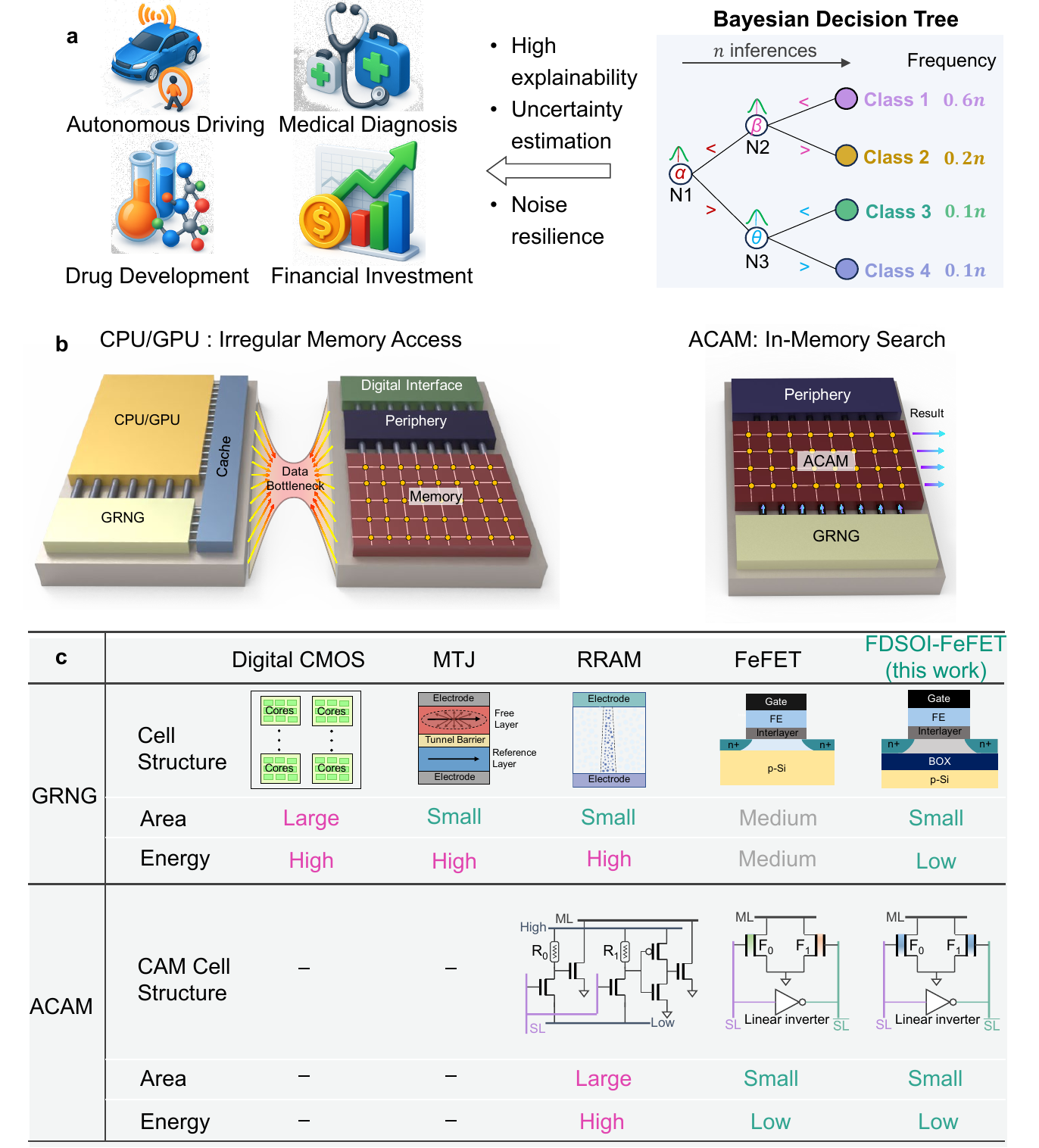}
    \caption{\textit{\textbf{ Motivation for FDSOI-FeFET-based robust BDT inference.} \textbf{a,} Applications such as autonomous driving, medical diagnosis, drug discovery and financial investment require machine learning models with strong interpretability, reliable uncertainty estimation and robustness to noise. BDT, in which each tree node is modeled as a Gaussian distribution and sampled during inference, inherently provides these properties by jointly enabling explainability, uncertainty estimation and noise resilience. \textbf{b,} Conventional implementations of BDTs rely on CPUs or GPUs for random number generation and probabilistic computation, with tree node parameters stored in memory. During inference, irregular memory access patterns and substantial data movement overhead are fundamentally limited by the von Neumann bottleneck. ACAM combined with efficient GRNG enables in-memory parallel search for BDT inference. \textbf{c,} Among candidate device technologies, FDSOI transistors with BTBT are well suited for energy-efficient GRNG, while FeFET provides compact cell area desirable for ACAM implementation. The proposed FDSOI–FeFET unified platform integrates stochastic generation and associative search within the same technology, offering a scalable and energy-efficient hardware solution for BDT implementation.}}
    \label{fig:motivation}
\end{figurehere}

 Nonetheless, the practical hardware realization of BDT presents considerable challenges. Two key building blocks of BDT are the tree inference engine and Gaussian Random Number Generator(GRNG) for threshold sampling. Both CPU- and GPU-based implementations suffer from the von Neumann bottleneck, where the separation of computation and memory limits data movement efficiency (Fig.~\ref{fig:motivation}b). Decision-tree inference is inherently serial, since the traversal of each node depends on the outcome of the previous one. Such a serial inference process results in irregular memory access patterns, further increasing data movement overhead and latency. Although GPUs offer parallel processing capabilities, their performance gains remain limited due to irregular accesses and thread workload imbalance \cite{xie2021}. Recent efforts to accelerate tree-based models leverage analog content-addressable memory (ACAM), which is hardware that compares input data with all the data stored in the ACAM array in parallel and outputs the address of the matched rows (Fig.~\ref{fig:motivation}c). Among the device candidates for ACAM, two main approaches have been widely explored: RRAM-based and FeFET-based implementations. 6T2R ACAM suffers from considerable variability and large area, which compromise scalibility and precision \cite{ielmini2018}. In contrast, FeFET-based ACAM is regarded as a state-of-the-art solution due to its efficient multi-bit storage capability and the smallest reported ACAM cell area \cite{yin2021,jerry2020,li2020}. 

For GRNG, several hardware platforms have also been investigated. Conventional CMOS-based analog GRNG typically yields limited randomness quality, restricting its applicability in high-security and precision-demanding scenarios \cite{Razavi2016}.  RRAM-based GRNG relies heavily on filament-forming variability, leading to inconsistency and degraded randomness stability, while also requiring high programming voltages that increase energy consumption \cite{yin2021,ielmini2018}.  The primary challenge of MRAM-based GRNG lies in the limited consistency of the generated random numbers and long-term stability, as the stochastic switching behavior is highly sensitive to process variations and environmental fluctuations, thereby degrading the reliability of GRNG \cite{khvalkovskiy2013}. Fully Depleted Silicon-On-Insulator (FDSOI) technology, providing a compact implementation and exploiting the band-to-band tunneling (BTBT) effect as an intrinsic entropy source, can be an attractive option as a high-quality solution for random number generation \cite{pei2025towards}.

Although ACAM and GRNG can be realized using separate device technologies, pursuing a unified single-technology solution is highly desirable for improved integration efficiency and reduce fabrication cost. However, employing one technology such as RRAM or FeFET to support both functionalities introduces notable trade-offs. RRAM suffers from inherent variability and large programming energy, degrading performance of ACAM and not well suited for sampling based application using GRNG. Conventional bulk FeFETs, although highly advantageous for ACAM, lack strong intrinsic entropy sources and therefore requires additional circuitry and power overhead to realize GRNG \cite{jerry2020,li2020}. 
In this context, FDSOI-FeFET emerges as a uniquely suitable technology for unified implementation of both ACAM and GRNG. 
The ferroelectric polarization hysteresis of FeFET enables robust, low-power, and high-density analog state storage for ACAM, while the BTBT effect occurring in the gate-to-drain overlap region of FDSOI structure under strong vertical electric fields and the floating body provides a high-quality entropy source for GRNG. Together, these features enable FDSOI-FeFETs as a highly integrated, area-efficient, and scalable hardware solution for BDT \cite{cesana2014,clermidy2016,yin2021}.

In this work, we propose a compact and energy-efficient hardware architecture for BDTs based on a single FDSOI-FeFET technology that natively supports both ACAM and GRNG. By exploiting device-level properties—ferroelectric polarization for multi-bit ACAM storage and BTBT for entropy generation, the proposed platform enables a dense, energy-efficient, and CMOS-compatible hardware implementation of probabilistic tree-based models. The major contributions of this work include: (i) proposing a single FDSOI FeFET-based technology that enables compact and energy efficient BDT execution, where multi-bit ACAM performs efficient decision-tree branch split while BTBT-induced entropy generation provides high-quality Gaussian random numbers required for probabilistic inference; (ii) applying software–hardware co-design to develop tailored mapping strategies and algorithmic optimizations for reducing the energy consumption and latency in BDT inference; (iii) demonstrating experimentally the key hardware primitives, including ACAM-based decision-tree branch execution and Gaussian random number generation implemented with FDSOI-FeFET devices; and (iv) conducting system-level evaluations to confirm the advantages of BDTs, where benchmarks on MNIST and a medical diagnosis dataset demonstrate reliable uncertainty estimation, high interpretability, and strong robustness to either noisy data or technology imperfections. These results highlight the potential of the proposed FDSOI-FeFET-based BDT inference accelerator as a scalable and practical solution for high-assurance AI in resource-constrained and safety-critical applications \cite{jerry2020,yin2021,pedretti2021,cesana2014,clermidy2016}.

\section*{\textcolor{sared}{\Large Results}}

\subsection*{Efficient Mapping Method}

As shown in Fig.~\ref{fig:mapping}, during inference of the BDT, the threshold at each tree node is first sampled from a customized Gaussian distribution. The input is then propagated through the tree to produce an output by following the path determined via comparing the input data with the  threshold values at the respective tree nodes. In subsequent iterations, a new threshold is independently sampled for each node, and the inference process is repeated. After n inference iterations, the final prediction is determined by selecting the class associated with the most frequently visited leaf node. To enable efficient hardware implementation of a BDT, the model first needs to be mapped onto the array composed of ACAM and GRNG. A key challenge arises from the requirement to sample random thresholds at each tree node during inference, which is initially difficult to implement directly in ACAM hardware. 

Conventional tree-to-ACAM mapping approaches adopt a feature-wise mapping strategy \cite{pedretti2021}, where each row corresponds to a root-to-leaf path in the tree and each column corresponds to a feature for area saving, as illustrated in Mapping-1 in Fig.\ref{fig:mapping}. Multiple tree nodes along different paths may share the same feature. During each inference, BDT requires sampling the threshold of each node from its customized Gaussian distribution. In this case, each column represents a feature, and since a feature may correspond to multiple tree nodes, the thresholds of the FeFET within a column can differ. Therefore, under this mapping strategy, to support randomly sampled node thresholds across inference iterations, it is necessary to reprogram each FeFET device within the ACAM cells in every inference iteration. After this step, deterministic feature input is applied across the whole column to determine the branch path. Such repeated programming incurs substantial energy and latency overheads and is further constrained by the limited endurance of FeFET devices.

To overcome these limitations, we propose a novel node-wise mapping strategy for BDT implementation, as illustrated as Mapping-2 in Fig.~\ref{fig:mapping}. In this scheme, each row of the ACAM corresponds to a decision path in the BDT, while each column represents a tree node. For columns corresponding to nodes that are not present in a given decision path, the associated ACAM cells are programmed to the don’t-care state, i.e., both FeFETs are programmed to the high-\textit{V}\textsubscript{TH} state so that they do not conduct current. Unlike conventional feature-wise mapping, the proposed approach incorporates threshold stochasticity into the input data rather than the CAM cell. This is motivated by observation that adding a random value to a node is the same as adding the random value to the input. Mapping each node to a column ensures that all cells within the same column share a common threshold value. Instead of programming a stochastic threshold into every cell, this shared threshold enables a simpler implementation: the mean threshold is programmed once into the corresponding CAM cells (deterministically, without sampling), while the randomly sampled component is combined with the original deterministic input through peripheral circuitry and applied as an input query for that column. In this way, during inference, random numbers generated by the GRNG are added to the query values of each column, effectively realizing random threshold sampling without modifying the programmed FeFET states. This design completely eliminates the need for reprogramming during inference, enabling a pure search-based operation. Consequently, the proposed mapping achieves significantly reduced energy consumption and latency, while effectively mitigating reliability and endurance constraints, making it well suited for efficient and scalable edge deployment of BDTs.

\begin{figurehere}
   \centering
    \includegraphics[scale=1,width=1\textwidth]{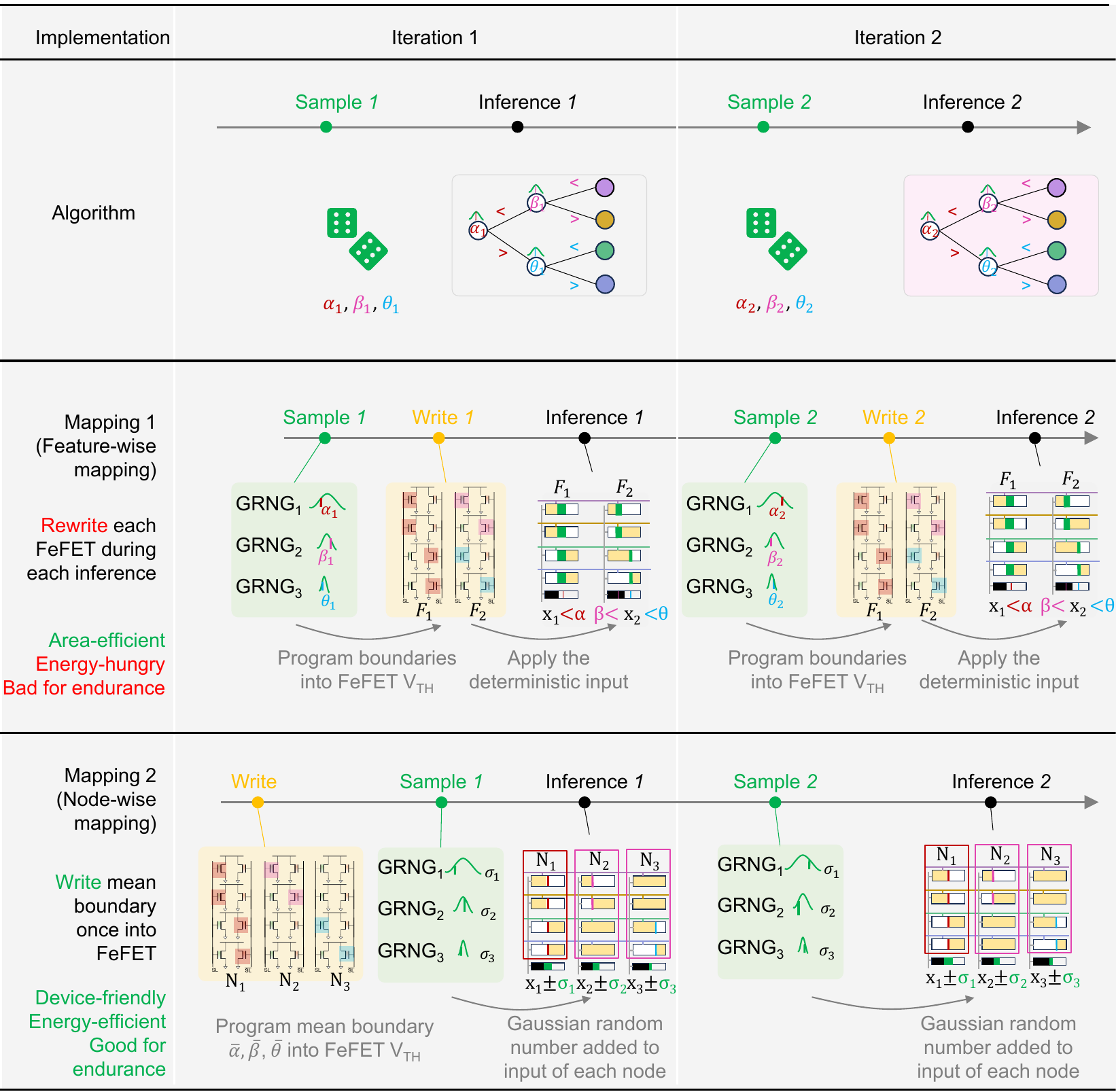}
    \caption{\textbf{Efficient method for mapping a BDT to ACAM.} \textit{During BDT inference, two operations are required in each iteration: Gaussian threshold sampling and tree inference. In a conventional feature-wise mapping (Mapping-1)), each column corresponds to a feature and each row represents a tree path. During the sampling phase, random thresholds generated by the GRNG is programmed into the threshold voltage of FeFET devices before searching with deterministic inputs. FeFET cells require reprogramming at every inference iteration. This mapping strategy is area-efficient, but repeated write operations incur substantial energy consumption and degrade device endurance. The proposed node-wise mapping (Mapping-2)), in which each tree node is mapped to an individual ACAM column. The mean threshold value is programmed once into the FeFET devices during initialization. During inference, Gaussian random numbers generated by the GRNG are directly added to the query voltage, enabling probabilistic inference without reprogramming the FeFET array. This approach eliminates per-iteration write operations, reducing energy consumption, improving inference latency, and enhancing device endurance.}} 
    \label{fig:mapping}
\end{figurehere}

\subsection*{FDSOI-FeFET GRNG}
To enable efficient random number generation, we introduce an FDSOI-FeFET device as the entropy source. As illustrated in Fig.~\ref{fig:grng}(a--c), the GRNG operation can be organized into a generation--read--reset cycle. During the generation phase, applying a negative gate bias together with a positive drain bias enhances the band bending of both the conduction and valence bands in the gate-to-drain overlap region. Under such a vertical electric field, electrons can tunnel across the forbidden bandgap through BTBT, thereby generating holes in the channel of the FDSOI-FeFET device. In the subsequent read phase, the accumulated holes modulate the channel conductivity and are reflected in the device read current. Since BTBT is inherently stochastic, the number of generated holes varies randomly from cycle to cycle, leading to fluctuations in the measured read current. This intrinsic current variability provides a high-quality entropy source for random number generation. During the reset phase, the stored holes can be erased by applying a positive gate bias together with a negative drain bias, restoring the device to its initial state and getting ready for the next cycle. This reversible write--erase behavior is conceptually analogous to the charge write and refresh mechanism in DRAM, while simultaneously enabling randomness extraction through stochastic hole generation. By leveraging this efficient generation--read--reset cycle within a single FDSOI-FeFET device, the proposed approach realizes highly compact and energy-efficient GRNG, offering significant advantages in both area and power consumption for scalable hardware integration.

\begin{figurehere}
   \centering
    \includegraphics[scale=1,width=1\textwidth]{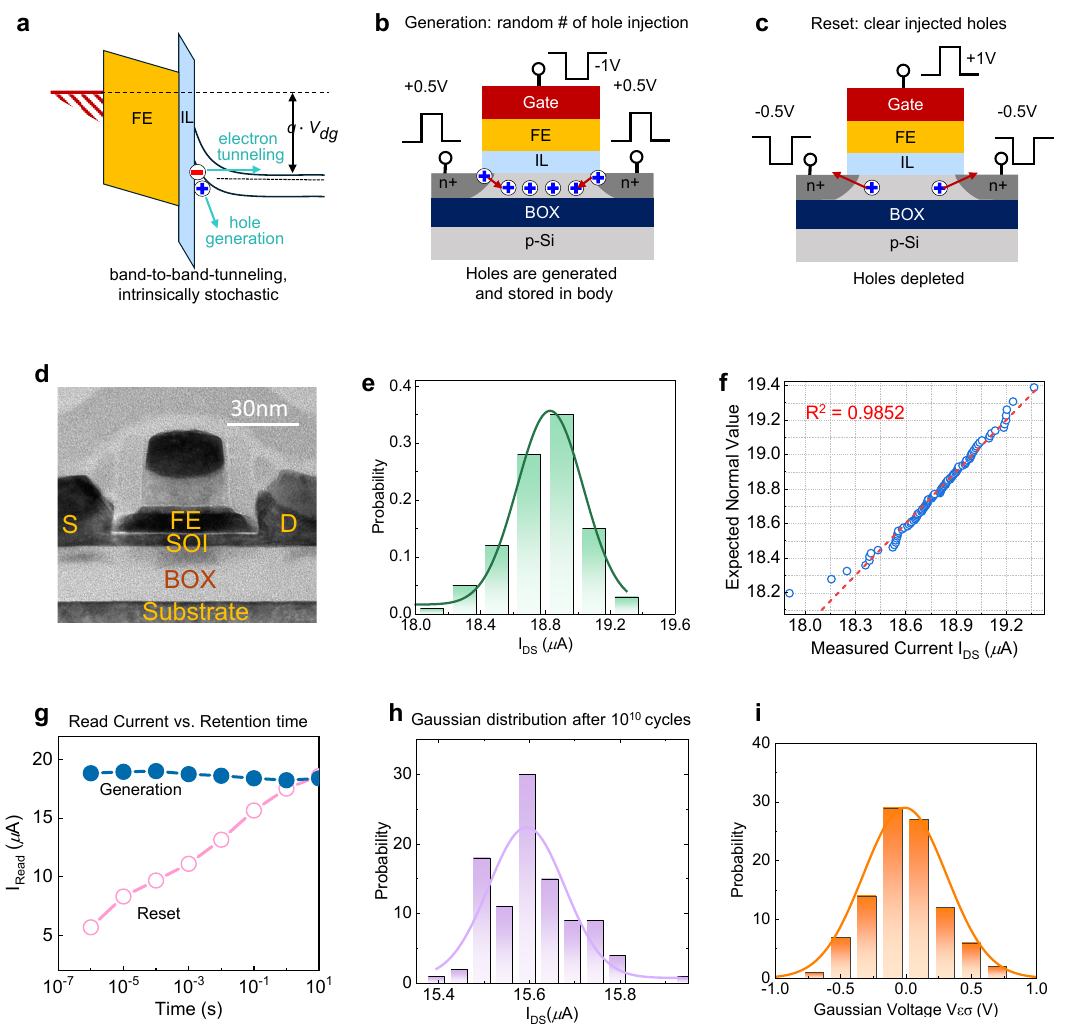}
    \caption{\textbf{GRNG Process and Reliability.} \textit{\textbf{a,} Negative gate bias in the FDSOI-FeFET induces enhanced band bending, enabling band-to-band tunnelling of electrons from the valence band to the conduction band and generating holes in the valence band. This process is intrinsically stochastic. \textbf{b,} During the generation phase, negative bias is applied to the gate while positive biases are applied to the drain and source, resulting in hole generation in the channel near the drain and source regions. \textbf{c,} During the reset phase, a positive gate bias and negative drain/source biases are applied, leading to hole depletion through the drain and source terminals. \textbf{d,} Cross-sectional transmission electron microscopy image of the device structure, showing the ferroelectric layer, silicon-on-insulator (SOI) layer and buried oxide layer. \textbf{e,} Distribution of 100 read currents after the generation phase, showing a Gaussian distribution. \textbf{f,} Quantile--quantile plot confirming the Gaussian-like distribution of the read current. \textbf{g,} Retention characteristics after the generation phase, demonstrating at least 1~ms retention, sufficient for GRNG operation. \textbf{h,} Endurance characteristics showing that, even after $10^{10}$ read--write cycles, the readout current maintains a Gaussian-like distribution. \textbf{i,} The Gaussian-distributed current is converted into a zero-mean Gaussian voltage using the circuit shown in the Supplementary Information.}} 
    \label{fig:grng}
\end{figurehere}

To validate the FDSOI based GRNG design, devices integrated on a 22nm FDSOI technology are adopted. Detailed fabrication details can be found in \cite{dunkel2017fefet}. A cross-sectional transmission electron microscopy (TEM) image of the fabricated device is shown in Fig.~\ref{fig:grng}d, clearly showing the ferroelectric layer, the silicon channel, and the buried oxide (BOX) layer. The statistical distribution of the sampled current of 100 cycles is shown in Fig.~\ref{fig:grng}e, while the corresponding Quantile–Quantile (QQ) Plot in Fig.~\ref{fig:grng}f confirms that the current obtained from 100 sampling cycles closely follows a Gaussian distribution. The device also demonstrates robust reliability. Fig.~\ref{fig:grng}g shows the retention behavior of the states after generation and reset. Since GRNG is not a memory operation, the reset state retention is not of interest here. The stable read current of state after generation over 1 ms indicates sufficient data holding time for sampling-based operation, while long-term retention is not required for BDT inference. After endurance cycling up to $10^{10}$ write--erase operations, the distribution of 100 sampled currents remains Gaussian-like, as shown in Fig.~\ref{fig:grng}h, though with a shift in the mean current value. To mitigate the effects of mean-value drift and device-to-device variation, a differential architecture is employed to generate a zero-mean Gaussian voltage. 

The complete GRNG circuit is shown in Fig.~\ref{fig:GRNG_circuit}. Two FDSOI transistors act as entropy sources that independently discharge the capacitors. Device variations in threshold voltage and discharge current lead to different discharge times ($T$), which are detected by a pair of inverters and an XOR gate, producing output pulses with Gaussian-distributed widths. The pulse width $T$ is then applied to the gate of a FeFET with programmable current (proportional to $\sigma_i$) to modulate the capacitor discharge time. This enables tuning of the distribution standard deviation $\sigma_i$ and shifting the mean value to zero by recharging the capacitors with compensation voltage $V_{\mathrm{COMP}}$. Finally, the Gaussian-distributed current is converted into a Gaussian voltage signal before being applied to the ACAM array (Fig.~\ref{fig:grng}i).

\subsection*{Experimental Demonstration of BDT}
As BDTs are well suited for uncertainty-critical and explainability-demanding scenarios, breast cancer diagnosis serves as an ideal application case for demonstration. To realize tree-branch splitting, it is necessary to implement both ``$>$'' and ``$<$'' comparison functions within a single ACAM cell. We experimentally demonstrate that programming a single FeFET at different branches of an ACAM cell can selectively shift the upper or lower boundary, which together define the matching region of the cell.  As shown in Fig.~\ref{fig:bdt}a, two FeFETs form one ACAM cell. The drains of the two FeFETs are connected to the same match line (ML), while their gates are driven complementarily through an inverter. Both sources are tied to ground. To configure the upper decision boundary, FeFET $F_1$ is programmed to a high-$V_{\mathrm{TH}}$ state, whereas FeFET $F_0$ is programmed to an analog threshold state that represents the desired upper boundary. To configure the lower decision boundary, FeFET $F_0$ is programmed to a high-$V_{\mathrm{TH}}$ state, whereas FeFET $F_1$ is programmed to an analog threshold state that represents the desired lower boundary. The measurement results show 8 levels of boundary storage of both lower boundary and upper boundary of 2FeFET ACAM for branch split. When search voltage falls outside the matching range, one of the FeFET is on and introduce a high current in ML. Fig.~\ref{fig:bdt}b shows the 3D matching range of an 1x2 ACAM array. The 3D colormap surface of the ML current and its projection onto the VSL$_1$--VSL$_2$ plane are presented. The results indicate that the low-current region (highlighted in green on the match plane) expands independently along each dimension as the $V_{\mathrm{TH}}$ of the $F_0$ transistor is adjusted in the corresponding ACAM cell. As shown in the 3D boundary visualization, the boundary expansion along both the VSL$_1$ and VSL$_2$ dimensions is achieved by programming $F_0$ to different analog $V_{\mathrm{TH}}$ states, thereby extending the upper boundary of the matching subspace. Inside this expanded ``basin'' region, the cell exhibits a low read current, whereas outside the boundary, the current rises sharply, indicating a mismatch condition. These results clearly demonstrate that each ACAM cell independently defines and controls the matching boundary along its associated dimension.

\begin{figurehere}
   \centering
    \includegraphics[scale=1,width=1\textwidth]{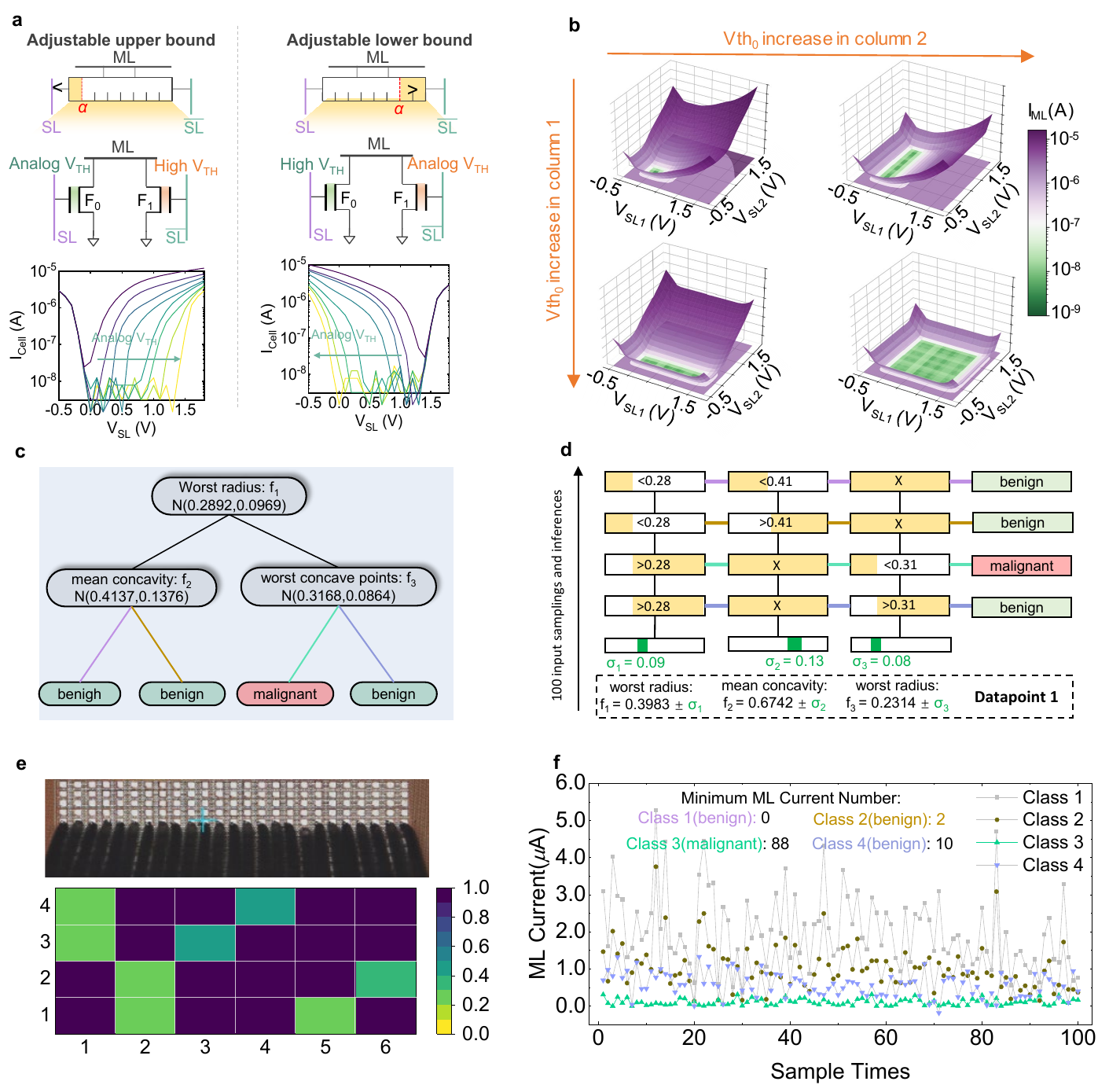}
    \caption{\textbf{BDT Inference Demonstration.}\textit{\textbf{a,} Implementation of ``$>$'' and ``$<$'' comparison operations in a two-FeFET ACAM cell by programming device threshold voltages to adjust the decision boundary. To tune the upper boundary, F1 is programmed to a high threshold voltage ($V_{\mathrm{th}}$) while F0 is set to an arbitrary analogue value. To tune the lower boundary, F0 is programmed to a high $V_{\mathrm{th}}$ and F1 to a low $V_{\mathrm{th}}$. Experimental results demonstrate eight discrete levels of boundary adjustment. \textbf{b,} Matching regions of a $1\times2$ ACAM array illustrates boundary adjustment across different feature dimensions. \textbf{c,} A two-layer decision tree trained on the breast cancer dataset. All node feature thresholds are modelled as Gaussian distributions and normalized prior to hardware mapping. \textbf{d,} ACAM array architecture showing the mapping of the BDT onto the memory array. \textbf{e,} Threshold voltage map of the $6\times4$ FeFET array implementing the mapped ACAM. \textbf{f,} Readout currents from different rows after 100 inference operations, corresponding to different classes. Final classification is determined using a winner-take-all scheme.}}
    \label{fig:bdt}
\end{figurehere}

After validating the branch-splitting functionality of a single ACAM cell, we demonstrate breast cancer diagnosis using a BDT mapped onto our ACAM array. Given the relatively low complexity of the dataset, a shallow decision tree with only 2–3 layers is sufficient for accurate classification. In this work, we employ a two-layer decision tree, which is mapped onto a 3×4 ACAM array, to demonstrate the feasibility of our approach. The diagnosis task determines whether a breast cancer sample is benign (non-cancerous) or malignant (cancerous) based on features extracted from digital microscope images in the Breast Cancer Wisconsin (Diagnostic) dataset. As shown in Fig.~\ref{fig:bdt}c, for demonstration, we employ a two-layer BDT consisting of three decision nodes. After training on the breast cancer dataset, the model automatically selects three representative features: \emph{worst radius}, \emph{mean concavity}, and \emph{worst area}, as the splitting dimensions. Through Bayesian training, both the selected features and the split thresholds of the decision nodes are learned from training data. Each node threshold is represented as a customized Gaussian distribution:
\begin{equation}T \sim \mathcal{N}(\mu,\,\sigma^{2}),\end{equation} where $\mu$ denotes the mean value of the learned threshold, and $\sigma$ captures the uncertainty of the decision boundary. The input features and the stored Gaussian parameters $(\mu,\sigma)$, as well as the FeFET \textit{V}\textsubscript{TH}, are normalized into the range of $[0,1]$. The mean value $\mu$ of each node threshold is directly programmed into the FeFET device within the corresponding ACAM cell by following  the approach discussed above. During inference, a random perturbation $\epsilon$ is generated by the GRNG at each column:\begin{equation}\epsilon \sim \mathcal{N}(0,\,\sigma^{2}),\end{equation}and added to the input feature value to enable on-the-fly sampling:\begin{equation}x' = x + \epsilon,\end{equation}, which effectively implements stochastic threshold sampling from the learned Gaussian distribution, as illustrated in Fig.~\ref{fig:bdt}d.

Fig.~\ref{fig:bdt}e shows the measurement setup and the experimentally readout threshold values after programming the ACAM array. The small programming error with a minor difference, smaller than 0.1 V, enables robust implementation of BDT. The device in unused cells are simply programmed to a high threshold (1.5V) for "don't care" state. The MLs of the ACAM array are first precharged to a high voltage to prepare for the inference. The input features combined with random numbers are then applied to the SLs of the array (where each feature is represented by a pair of SLs connected via an analog inverter), causing the MLs to discharge according to how well the input matches each path’s conditions. After 100 inference iterations, we obtain the ML current corresponding to each decision path in every sampled tree realization. For datapoint~1, the third path exhibits the smallest ML current in 88 out of 100 inferences, indicating that this path is selected with an 88\% frequency. Therefore, the datapoint can be classified as malignant with an estimated confidence of 88\% (Fig.\ref{fig:bdt}f).

\subsection*{Variation Robustness of BDT}
One of the key advantages of BDTs is their strong noise resilience enabled by sampling-based inference. This allows BDTs to maintain high classification accuracy even when query data is noisy, where conventional deterministic models may suffer significant performance degradation. Meanwhile, emerging memory-based ACAM technologies, such as FeFET and RRAM, face additional challenges compared with mature CMOS implementations. The stored thresholds in FeFET-based devices may deviate from their intended values because of non-idealities in read and write operations due to process variations or intrinsic switching stochasticity. BDTs can effectively mitigate this issue by sampling the decision thresholds during inference and averaging across samples, thereby reducing the impact of noise and improving overall robustness. 

To evaluate BDT's noise resilience, we conduct simulations on the MNIST dataset. The ACAM implementation is modeled using the CAM simulation tool \textit{CAMASim} \cite{li2024camasim}. After training, both conventional DTs and BDTs are mapped onto the ACAM for inference. To evaluate robustness against input noise, we generate noisy datasets by injecting zero-mean Gaussian noise into the normalized MNIST test datasets. The noise magnitude is controlled by the standard deviation~$\sigma$, and the resulting classification accuracy is measured on the test set. To assess tolerance to device-level variations, we introduce perturbations to the stored thresholds of individual FeFET devices in the ACAM, emulating the variation during read operation in practical hardware.The threshold voltage is programmed according to the mapping method in Fig.~\ref{fig:mapping}. During inference, noise of different magnitudes is added to the FeFET threshold voltage in \textit{CAMASim}, which is a modular and extensible simulation framework that takes search-centric applications written in Python as input and emulates search behavior to generate match outcomes and estimate overall application-level accuracy. Because the tree node threshold is normalized during mapping, the noise magnitude directly corresponds to the voltage perturbation. For example, 10\% noise corresponds to a 0.1\,V variation in $V_{\mathrm{th}}$. The FDSOI-FeFET model in \textit{CAMASim} is calibrated with the experimental measurement data, the fitting curve and measured IV of the device is shown in Fig.~\ref{fig:model}.

\begin{figurehere}
   \centering
    \includegraphics[scale=1,width=0.8\textwidth]{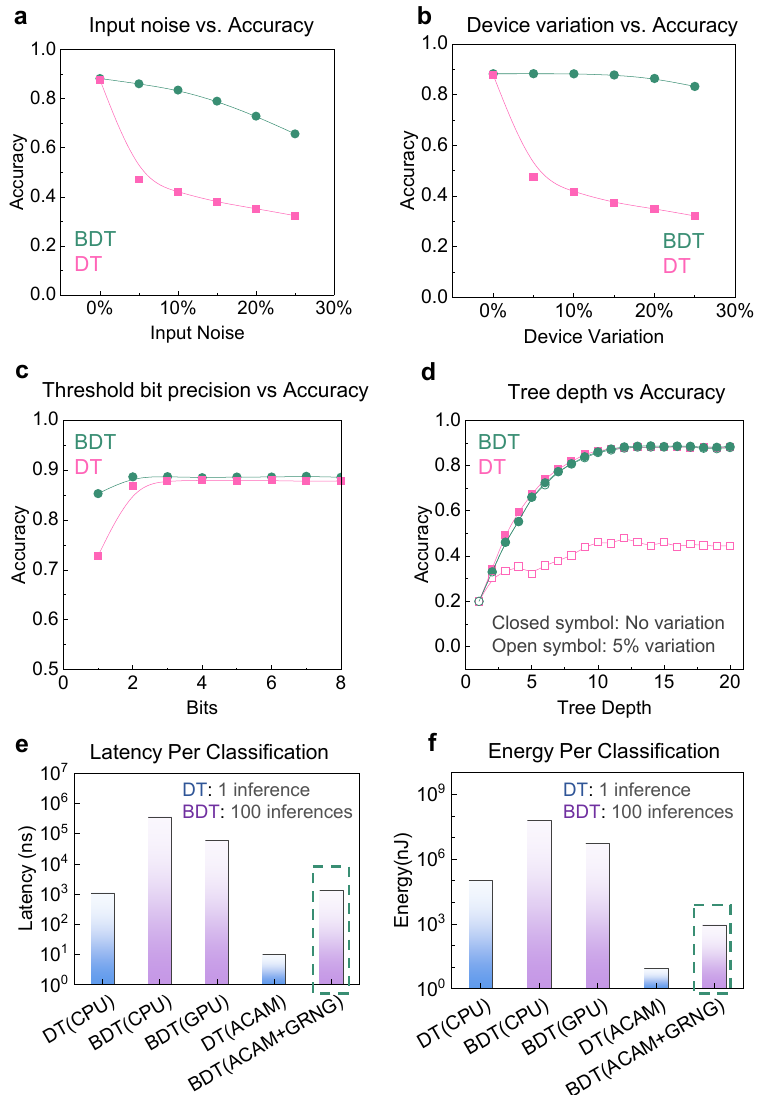}
    \caption{\textbf{Variation Robustness of BDT.} \textit{\textbf{a,} Input noise resilience evaluated on the MNIST dataset. All input data are normalized to the range [0,1], and additive Gaussian noise is introduced to each data point, with values clipped to remain within [0,1]. The results demonstrate strong robustness against input perturbations. \textbf{b,} Device-level variation analysis using CAMASim. Threshold voltage ($V_{\mathrm{th}}$) variation is independently added to each FeFET, with $V_{\mathrm{th}}$ distributed within [0,1]. Query voltages are also normalized to [0,1]. The results indicate robustness to device variability. \textbf{c,} Simulation results showing classification accuracy as a function of threshold quantization bits, indicating that 2-bit precision is sufficient for MNIST classification. \textbf{d,} Noise resilience under different tree depths, demonstrating that the BDT maintains robustness even for deeper tree architectures. \textbf{e,} Latency per classification, showing approximately two orders of magnitude reduction compared with a GPU implementation. \textbf{f,} Energy consumption per classification, showing approximately four orders of magnitude reduction compared with a GPU implementation.}}
    \label{fig:robustness}
\end{figurehere}

As shown in Fig.~\ref{fig:robustness}a and Fig.~\ref{fig:robustness}b, the BDT consistently demonstrates superior robustness to both input noise and device variations compared to the conventional DT. Since FDSOI-FeFET devices can only support limited bit precision,, we further investigate the required precision for storing tree node thresholds. The results in Fig.~\ref{fig:robustness}c indicate that a 2-bit representation is sufficient to achieve accuracy comparable to give precise implementations on the MNIST dataset. We additionally evaluate the noise resilience of DT and BDT models with increasing tree depth. As illustrated in Fig.~\ref{fig:robustness}d, even at larger depths, the BDT maintains strong noise tolerance and is able to recover baseline accuracy levels observed under noise-free conditions. Fig.~\ref{fig:robustness}e and Fig.~\ref{fig:robustness}f present the inference latency and energy consumption of DT and BDT models at a tree depth of 20. At this depth, both models share an identical topology consisting of more than 3000 root-to-leaf paths. 

Despite its higher computational complexity, BDT inference on the ACAM architecture achieves substantial acceleration over CPU (Intel Core i9-14900) and GPU (RTX-4060) baselines, delivering a speedup of approximately 2–3 orders of magnitude. Moreover, the energy efficiency is significantly improved, with a reduction of about 4–5 orders of magnitude compared to conventional CPU and GPU implementations. The benchmark results are summarized in Table~\ref{tab:latency_energy}.

\section*{\textcolor{sared}{\Large Discussion}}

This work presents a hardware-algorithm co-design for accelerating BDT inference. It demonstrates that BDT can be effectively accelerated using analog in-memory computing when co-designed with device and architectural properties. By leveraging a single FDSOI-FeFET technology, the proposed architecture integrates ACAM and GRNG to enable efficient, explainable, and uncertainty-aware inference without incurring reprogramming overhead. The node-wise mapping strategy shifts stochasticity from memory states to the input domain, avoiding repeated device programming that may degrade device reliability, while improving inference speed and reducing energy consumption. As a result, BDT inference exhibits strong robustness to both noisy data and hardware variations, while achieving orders-of-magnitude improvements in latency and energy efficiency compared to conventional CPU or GPU implementations. These results suggest that probabilistic and interpretable models, when aligned with emerging device physics, offer a promising pathway toward scalable and trustworthy artificial intelligence in resource-constrained and safety-critical systems.

\section*{\textcolor{sared}{\Large Methods}}

\subsection*{FDSOI FeFET Device fabrication}
The FDSOI FeFETs were fabricated at GlobalFoundries using the 22nm technology node. The device features a stack composed of poly-crystalline Si/TiN/doped HfO\textsubscript{2}/SiO\textsubscript{2}/Si/BOX/substrate. The buried oxide is 20nm SiO\textsubscript{2}. Detailed information
can be found in \cite{dunkel2017fefet}. The ferroelectric gate stack process module starts with growth of a thin
SiO\textsubscript{2} based interfacial layer, followed by the deposition of the doped HfO\textsubscript{2} film via atomic
layer deposition (ALD). A TiN metal gate electrode was deposited using physical vapor
deposition, on top of which the poly-Si gate electrode is deposited. The source
and drain doped regions were then activated by a rapid thermal annealing at approximately 1000 $^\circ$C. This step also results in the formation of the ferroelectric orthorhombic
phase within the doped HfO\textsubscript{2}.

\subsection*{Gaussian Random Voltage Generation}
Gaussian random voltage is generated from the Gaussian read-out current. 
In Fig.~\ref{fig:grng}, we introduced how a random current is generated. 
During the query operation in the ACAM array, a Gaussian random voltage is applied to the input voltage. 
As shown in the supplementary material, the first step is to generate two independent random currents from two different FDSOI devices. 
These currents charge two capacitors in two separate branches and discharge at different rates to generate a Gaussian pulse:

\begin{equation}
T \sim \mathcal{N}(\mu_P, \sigma_P^2)
\end{equation}

The scaling factor

\begin{equation}
\sigma' = \frac{\sigma}{\sigma_P}
\end{equation}

is stored as the threshold voltage of the FeFET. 
The signal $T$ controls the FeFET gate with a programmable current (proportional to $\sigma'$), which fine-tunes the distribution to

\begin{equation}
\sigma' T 
\sim \mathcal{N}
\left(
\frac{\sigma}{\sigma_P}\mu_P, \sigma^2
\right)
\end{equation}

Afterward, the Gaussian voltage is added to a compensation voltage stored in a capacitor to generate a zero-mean Gaussian voltage:

\begin{equation}
V_{\epsilon\sigma}
= \sigma' T - \frac{\sigma}{\sigma_P}\mu_P
\sim \mathcal{N}(0, \sigma^2)
\end{equation}

Finally, the zero-mean Gaussian voltage $V_{\epsilon\sigma}$ is added to the query voltage $V_X$ during BDT inference.

\subsection*{Training algorithm of tree models}
All hard tree-based models are trained in a Python environment with Scikit-learn package. BDT is trained using a probabilistic threshold selection framework derived from impurity reduction statistics. 
At each node, the impurity of the label set $\bm{y}$ is quantified using the Gini index,

\begin{equation}
G(\bm{y}) 
= 1 - \sum_{k=1}^{K} 
\left(
\frac{\mathrm{count}(y = k)}{\bm{y}}
\right)^2 ,
\end{equation}

where $K$ is the number of classes and $\bm{y}$ denotes the number of samples at the current node. 
For each candidate feature $f$ and threshold $t$, the dataset is partitioned into two subsets according to $x_{i,f} \le t$ and $x_{i,f} > t$. 
The resulting weighted impurity reduction is defined as

\begin{equation}
\Delta G(t) 
= G(\bm{y}) 
- \frac{\bm{y}_{\mathrm{left}}}{\bm{y}} G(\bm{y}_{\mathrm{left}})
- \frac{\bm{y}_{\mathrm{right}}}{\bm{y}} G(\bm{y}_{\mathrm{right}}).
\end{equation}

Rather than deterministically selecting the threshold that maximizes $\Delta G(t)$, we construct a probability distribution over all valid thresholds for a given feature. 
Each threshold is assigned a non-negative weight,

\begin{equation}
w(t) = \max(\Delta G(t), 0),
\end{equation}

which is normalized to obtain a discrete probability distribution,

\begin{equation}
p(t) = 
\frac{w(t)}
{\sum_{t'} w(t')}.
\end{equation}

The mean and variance of the threshold distribution are then computed as

\begin{align}
\mu_f &= \sum_{t} p(t)\, t, \\
\sigma_f^2 &= \sum_{t} p(t)\, (t - \mu_f)^2.
\end{align}

Finally, the threshold associated with feature $f$ is modeled as a Gaussian random variable,

\begin{equation}
t_f \sim \mathcal{N}(\mu_f, \sigma_f^2).
\end{equation}

This probabilistic representation enables direct compatibility with the hardware implementation, in which Gaussian perturbations are physically generated and injected during inference. 
The learned mean $\mu_f$ defines the nominal comparison voltage and is pre-programmed to the FDSOI-FeFET ACAM array, while $\sigma_f$ determines the stochastic amplitude realized through FDSOI-FeFET BTBT scheme.

\bibliography{ref.bib}

\begin{thebibliography}{1}
\expandafter\ifx\csname url\endcsname\relax
  \def\url#1{\texttt{#1}}\fi
\expandafter\ifx\csname urlprefix\endcsname\relax\def\urlprefix{URL }\fi
\providecommand{\bibinfo}[2]{#2}
\providecommand{\eprint}[2][]{\url{#2}}

\bibitem{intel7_wiki}
\bibinfo{author}{{Nenni, D.}}
\newblock \bibinfo{title}{Intel 7 process technology}.
\newblock \bibinfo{howpublished}{\url{https://semiwiki.com/wikis/industry-wikis/intel-7-process-technology-wiki/}} (\bibinfo{year}{2023}).

\bibitem{ada_lovelace}
\bibinfo{author}{{NVIDIA Corporation}}.
\newblock \bibinfo{title}{Ada lovelace gpu architecture}.
\newblock \bibinfo{howpublished}{\url{https://en.wikipedia.org/wiki/Ada_Lovelace_(microarchitecture)}} (\bibinfo{year}{2022}).

\bibitem{intel_i9_2024}
\bibinfo{author}{{Intel Corporation}}.
\newblock \bibinfo{title}{Intel core i9 processors (14th gen)}.
\newblock \bibinfo{howpublished}{\url{https://www.intel.com/content/www/us/en/ark/products/series/236143/intel-core-i9-processors-14th-gen.html}} (\bibinfo{year}{2024}).

\bibitem{nvidia_4060_2023}
\bibinfo{author}{{NVIDIA Corporation}}.
\newblock \bibinfo{title}{Geforce rtx 4060 graphics card}.
\newblock \bibinfo{howpublished}{\url{https://www.nvidia.com/en-us/geforce/graphics-cards/40-series/rtx-4060-4060ti/}} (\bibinfo{year}{2023}).

\end{thebibliography}


\begin{thebibliography}{10}
\expandafter\ifx\csname url\endcsname\relax
  \def\url#1{\texttt{#1}}\fi
\expandafter\ifx\csname urlprefix\endcsname\relax\def\urlprefix{URL }\fi
\providecommand{\bibinfo}[2]{#2}
\providecommand{\eprint}[2][]{\url{#2}}

\bibitem{grigorescu2020}
\bibinfo{author}{Grigorescu, S.}, \bibinfo{author}{Trasnea, B.}, \bibinfo{author}{Cocias, T.} \& \bibinfo{author}{Macesanu, G.}
\newblock \bibinfo{title}{A survey of deep learning techniques for autonomous driving}.
\newblock \emph{\bibinfo{journal}{Journal of Field Robotics}} \textbf{\bibinfo{volume}{37}}, \bibinfo{pages}{362--386} (\bibinfo{year}{2020}).

\bibitem{feng2021}
\bibinfo{author}{Feng, D.}, \bibinfo{author}{Haase-Schütz, C.}, \bibinfo{author}{Rosenbaum, L.}, \bibinfo{author}{Hertlein, H.}, \bibinfo{author}{Gläser, C.}, \bibinfo{author}{Timm, F.} \& \bibinfo{author}{Dietmayer, K.}
\newblock \bibinfo{title}{Deep multi-modal object detection and semantic segmentation for autonomous driving: Datasets, methods, and challenges}.
\newblock \emph{\bibinfo{journal}{IEEE Transactions on Intelligent Transportation Systems}} \textbf{\bibinfo{volume}{22}}, \bibinfo{pages}{1341--1360} (\bibinfo{year}{2021}).

\bibitem{esteva2019}
\bibinfo{author}{Esteva, A. e.~a.}
\newblock \bibinfo{title}{A guide to deep learning in healthcare}.
\newblock \emph{\bibinfo{journal}{Nature Medicine}} \textbf{\bibinfo{volume}{25}}, \bibinfo{pages}{24--29} (\bibinfo{year}{2019}).

\bibitem{rajpurkar2022}
\bibinfo{author}{Rajpurkar, P.}, \bibinfo{author}{Chen, E.}, \bibinfo{author}{Banerjee, O.} \& \bibinfo{author}{Topol, E.~J.}
\newblock \bibinfo{title}{Ai in health and medicine}.
\newblock \emph{\bibinfo{journal}{Nature Medicine}} \textbf{\bibinfo{volume}{28}}, \bibinfo{pages}{31--38} (\bibinfo{year}{2022}).

\bibitem{dixon2020}
\bibinfo{author}{Dixon, M.~F.}, \bibinfo{author}{Halperin, I.} \& \bibinfo{author}{Bilokon, P.}
\newblock \emph{\bibinfo{title}{Machine Learning in Finance: From Theory to Practice}} (\bibinfo{publisher}{Springer International Publishing}, \bibinfo{year}{2020}).

\bibitem{fischer2018}
\bibinfo{author}{Fischer, T.} \& \bibinfo{author}{Krauss, C.}
\newblock \bibinfo{title}{Deep learning with long short-term memory networks for financial market predictions}.
\newblock \emph{\bibinfo{journal}{European Journal of Operational Research}} \textbf{\bibinfo{volume}{270}}, \bibinfo{pages}{654--669} (\bibinfo{year}{2018}).

\bibitem{bonnet2023bringing}
\bibinfo{author}{Bonnet, D.}, \bibinfo{author}{Hirtzlin, T.}, \bibinfo{author}{Majumdar, A.}, \bibinfo{author}{Dalgaty, T.}, \bibinfo{author}{Esmanhotto, E.}, \bibinfo{author}{Meli, V.}, \bibinfo{author}{Castellani, N.}, \bibinfo{author}{Martin, S.}, \bibinfo{author}{Nodin, J.-F.}, \bibinfo{author}{Bourgeois, G.} \emph{et~al.}
\newblock \bibinfo{title}{Bringing uncertainty quantification to the extreme-edge with memristor-based bayesian neural networks}.
\newblock \emph{\bibinfo{journal}{Nature Communications}} \textbf{\bibinfo{volume}{14}}, \bibinfo{pages}{7530} (\bibinfo{year}{2023}).

\bibitem{esteva2017dermatologist}
\bibinfo{author}{Esteva, A.}, \bibinfo{author}{Kuprel, B.}, \bibinfo{author}{Novoa, R.~A.}, \bibinfo{author}{Ko, J.}, \bibinfo{author}{Swetter, S.~M.}, \bibinfo{author}{Blau, H.~M.} \& \bibinfo{author}{Thrun, S.}
\newblock \bibinfo{title}{Dermatologist-level classification of skin cancer with deep neural networks}.
\newblock \emph{\bibinfo{journal}{Nature}} \textbf{\bibinfo{volume}{542}}, \bibinfo{pages}{115--118} (\bibinfo{year}{2017}).

\bibitem{lundberg2020local}
\bibinfo{author}{Lundberg, S.~M.}, \bibinfo{author}{Erion, G.}, \bibinfo{author}{Chen, H.}, \bibinfo{author}{DeGrave, A.}, \bibinfo{author}{Prutkin, J.~M.}, \bibinfo{author}{Nair, B.} \emph{et~al.}
\newblock \bibinfo{title}{From local explanations to global understanding with explainable ai for trees}.
\newblock \emph{\bibinfo{journal}{Nature Machine Intelligence}} \textbf{\bibinfo{volume}{2}}, \bibinfo{pages}{56--67} (\bibinfo{year}{2020}).

\bibitem{pedretti2021}
\bibinfo{author}{Pedretti, G. e.~a.}
\newblock \bibinfo{title}{Tree-based machine learning performed in-memory with memristive analog cam}.
\newblock \emph{\bibinfo{journal}{Nature Communications}} \textbf{\bibinfo{volume}{12}}, \bibinfo{pages}{5806} (\bibinfo{year}{2021}).

\bibitem{nakahara2025}
\bibinfo{author}{Nakahara, Y. e.~a.}
\newblock \bibinfo{title}{Bayesian decision theory on decision trees: Uncertainty evaluation and interpretability}.
\newblock In \emph{\bibinfo{booktitle}{Proceedings of the 28th International Conference on Artificial Intelligence and Statistics}}, vol. \bibinfo{volume}{258}, \bibinfo{pages}{1045--1053} (\bibinfo{year}{2025}).

\bibitem{lin2023uncertainty}
\bibinfo{author}{Lin, Y.}, \bibinfo{author}{Zhang, Q.}, \bibinfo{author}{Gao, B.}, \bibinfo{author}{Tang, J.}, \bibinfo{author}{Yao, P.}, \bibinfo{author}{Li, C.}, \bibinfo{author}{Huang, S.}, \bibinfo{author}{Liu, Z.}, \bibinfo{author}{Zhou, Y.}, \bibinfo{author}{Liu, Y.} \emph{et~al.}
\newblock \bibinfo{title}{Uncertainty quantification via a memristor bayesian deep neural network for risk-sensitive reinforcement learning}.
\newblock \emph{\bibinfo{journal}{Nature Machine Intelligence}} \textbf{\bibinfo{volume}{5}}, \bibinfo{pages}{714--723} (\bibinfo{year}{2023}).

\bibitem{nuti2021}
\bibinfo{author}{Nuti, G.}, \bibinfo{author}{Jiménez~Rugama, L.~A.} \& \bibinfo{author}{Cross, A.-I.}
\newblock \bibinfo{title}{An explainable bayesian decision tree algorithm}.
\newblock \emph{\bibinfo{journal}{Frontiers in Applied Mathematics and Statistics}} \textbf{\bibinfo{volume}{7}}, \bibinfo{pages}{598833} (\bibinfo{year}{2021}).

\bibitem{xie2021}
\bibinfo{author}{Xie, Z.}, \bibinfo{author}{Dong, W.}, \bibinfo{author}{Liu, J.}, \bibinfo{author}{Liu, H.} \& \bibinfo{author}{Li, D.}
\newblock \bibinfo{title}{Tahoe: Tree structure-aware high performance inference engine for decision tree ensemble on gpu}.
\newblock In \emph{\bibinfo{booktitle}{Proceedings of the 16th European Conference on Computer Systems (EuroSys)}}, \bibinfo{pages}{386--401} (\bibinfo{year}{2021}).

\bibitem{ielmini2018}
\bibinfo{author}{Ielmini, D.} \& \bibinfo{author}{Wong, H.-S.~P.}
\newblock \bibinfo{title}{In-memory computing with resistive switching devices}.
\newblock \emph{\bibinfo{journal}{Nature Electronics}} \textbf{\bibinfo{volume}{1}}, \bibinfo{pages}{333--343} (\bibinfo{year}{2018}).

\bibitem{yin2021}
\bibinfo{author}{Yin, X.}, \bibinfo{author}{M{\"u}ller, F.}, \bibinfo{author}{Laguna, A.~F.}, \bibinfo{author}{Li, C.}, \bibinfo{author}{Huang, Q.}, \bibinfo{author}{Shi, Z.}, \bibinfo{author}{Lederer, M.}, \bibinfo{author}{Laleni, N.}, \bibinfo{author}{Deng, S.}, \bibinfo{author}{Zhao, Z.} \emph{et~al.}
\newblock \bibinfo{title}{Deep random forest with ferroelectric analog content addressable memory}.
\newblock \emph{\bibinfo{journal}{Science advances}} \textbf{\bibinfo{volume}{10}}, \bibinfo{pages}{eadk8471} (\bibinfo{year}{2024}).

\bibitem{jerry2020}
\bibinfo{author}{Jerry, M. e.~a.}
\newblock \bibinfo{title}{Ferroelectric fet analog synapse for acceleration of deep neural network training}.
\newblock \emph{\bibinfo{journal}{IEEE Transactions on Electron Devices}} \textbf{\bibinfo{volume}{67}}, \bibinfo{pages}{667--674} (\bibinfo{year}{2020}).

\bibitem{li2020}
\bibinfo{author}{Li, Y.}, \bibinfo{author}{Luo, H.}, \bibinfo{author}{Zhang, X.}, \bibinfo{author}{He, Q.} \& \bibinfo{author}{Sun, Y.}
\newblock \bibinfo{title}{Ferroelectric field-effect transistors for memory and computing}.
\newblock \emph{\bibinfo{journal}{Journal of Semiconductors}} \textbf{\bibinfo{volume}{41}}, \bibinfo{pages}{021101} (\bibinfo{year}{2020}).

\bibitem{Razavi2016}
\bibinfo{author}{Razavi, B.}
\newblock \emph{\bibinfo{title}{Design of Analog CMOS Integrated Circuits}} (\bibinfo{publisher}{McGraw-Hill Education}, \bibinfo{year}{2016}).

\bibitem{khvalkovskiy2013}
\bibinfo{author}{Khvalkovskiy, A.}, \bibinfo{author}{Apalkov, D.} \& \bibinfo{author}{Watts, S. e.~a.}
\newblock \bibinfo{title}{Basic principles of stt-mram cell operation in memory arrays}.
\newblock \emph{\bibinfo{journal}{Journal of Physics D: Applied Physics}} \textbf{\bibinfo{volume}{46}}, \bibinfo{pages}{074001} (\bibinfo{year}{2013}).

\bibitem{pei2025towards}
\bibinfo{author}{Pei, L.}, \bibinfo{author}{Zhou, Y.}, \bibinfo{author}{Wang, X.}, \bibinfo{author}{Zhao, X.}, \bibinfo{author}{Huang, W.}, \bibinfo{author}{Cheng, B.}, \bibinfo{author}{Mulaosmanovic, H.}, \bibinfo{author}{Duenkel, S.}, \bibinfo{author}{Kleimaier, D.}, \bibinfo{author}{Beyer, S.} \emph{et~al.}
\newblock \bibinfo{title}{Towards uncertainty-aware robotic perception via mixed-signal bnn engine leveraging probabilistic quantum tunneling}.
\newblock In \emph{\bibinfo{booktitle}{2025 62nd ACM/IEEE Design Automation Conference (DAC)}}, \bibinfo{pages}{1--7} (\bibinfo{organization}{IEEE}, \bibinfo{year}{2025}).

\bibitem{cesana2014}
\bibinfo{author}{Cesana, G. e.~a.}
\newblock \bibinfo{title}{Low power design using fdsoi technology}.
\newblock In \emph{\bibinfo{booktitle}{MPSOC Forum}} (\bibinfo{year}{2014}).

\bibitem{clermidy2016}
\bibinfo{author}{Clermidy, F.}
\newblock \bibinfo{title}{Fdsoi technology general overview}.
\newblock In \emph{\bibinfo{booktitle}{SITRI}} (\bibinfo{year}{2016}).

\bibitem{dunkel2017fefet}
\bibinfo{author}{D{\"u}nkel, S.}, \bibinfo{author}{Trentzsch, M.}, \bibinfo{author}{Richter, R.}, \bibinfo{author}{Moll, P.}, \bibinfo{author}{Fuchs, C.}, \bibinfo{author}{Gehring, O.}, \bibinfo{author}{Majer, M.}, \bibinfo{author}{Wittek, S.}, \bibinfo{author}{M{\"u}ller, B.}, \bibinfo{author}{Melde, T.} \emph{et~al.}
\newblock \bibinfo{title}{A fefet based super-low-power ultra-fast embedded nvm technology for 22nm fdsoi and beyond}.
\newblock In \emph{\bibinfo{booktitle}{2017 IEEE International Electron Devices Meeting (IEDM)}}, \bibinfo{pages}{19--7} (\bibinfo{organization}{IEEE}, \bibinfo{year}{2017}).

\bibitem{li2024camasim}
\bibinfo{author}{Li, M.}, \bibinfo{author}{Liu, S.}, \bibinfo{author}{Sharifi, M.~M.} \& \bibinfo{author}{Hu, X.~S.}
\newblock \bibinfo{title}{{CAMASim}: A comprehensive simulation framework for content-addressable memory based accelerators}.
\newblock \emph{\bibinfo{journal}{arXiv preprint arXiv:2403.03442}}  (\bibinfo{year}{2024}).

\end{thebibliography}

\bibliographystyle{naturemag}

\section*{\large Acknowledgments}

This work is primarily supported by NSF 2346953, 2347024, 2235472, and 2404874. The experimental characterization of the bayesian decision tree is supported by SUPREME center, one of the SRC/DARPA JUMP 2.0 centers. 
The testing chips are partially funded by the European Union within "NextGeneration EU", by the Federal Ministry for Economic Affairs and Energy (BMWE) on the basis of a decision by the German Bundestag and by the State of Saxony with tax revenues based on the budget approved by the members of the Saxon State Parliament in the framework of “Important Project of Common European Interest - Microelectronics and Communication Technologies", under the project name “EUROFOUNDRY”.

\section*{\large Author contributions}

X.S.H., K.N. and N.C. proposed the project. P.R. led the project. P.R., X.W., J.D., and X.N. performed the experiments and collected the device characterization data. P.R., B.C., and N.C. completed the circuit design and simulations. H.M., S.D., and S.B. fabricated the devices. P.R. carried out the algorithm design and simulations. X.S.H., K.N. and N.C. contributed to technical discussions. All authors contributed to the writing of the manuscript.

\section*{\large Competing interests}
The authors declare that they have no competing interests.

\newpage

\renewcommand{\thefigure}{S\arabic{figure}}
\renewcommand{\thetable}{S\arabic{table}}
\renewcommand{\theequation}{S\arabic{equation}}
\setcounter{figure}{0}
\setcounter{table}{0}
\setcounter{equation}{0}
\begin{bibunit}
\newpage
\begin{center}
\title{\textbf{\Large Supplementary Materials}}
\end{center}

\begin{figurehere}
   \centering
    \includegraphics[scale=1,width=1\textwidth]{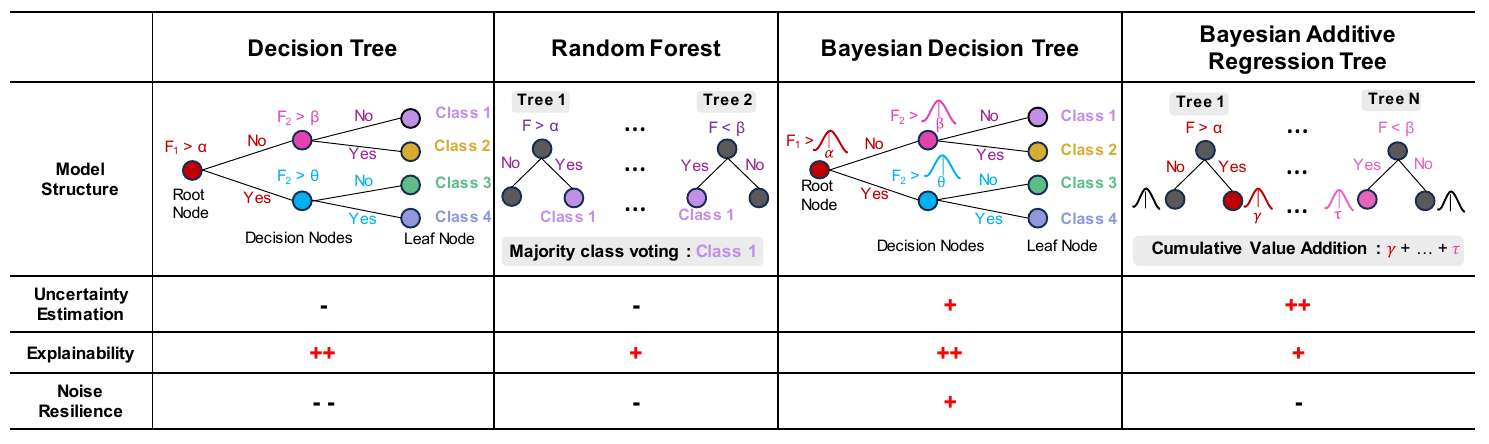}
    \caption{\textbf{Different tree models comparison.} Uncertainty estimation, explainability, and noise resilience comparison across mainstream tree models shows that BDT is the one that simultaneously satisfies all three requirements.}
    \label{fig:comparision}
\end{figurehere}

\newpage
\begin{figurehere}
   \centering
    \includegraphics[scale=1,width=1\textwidth]{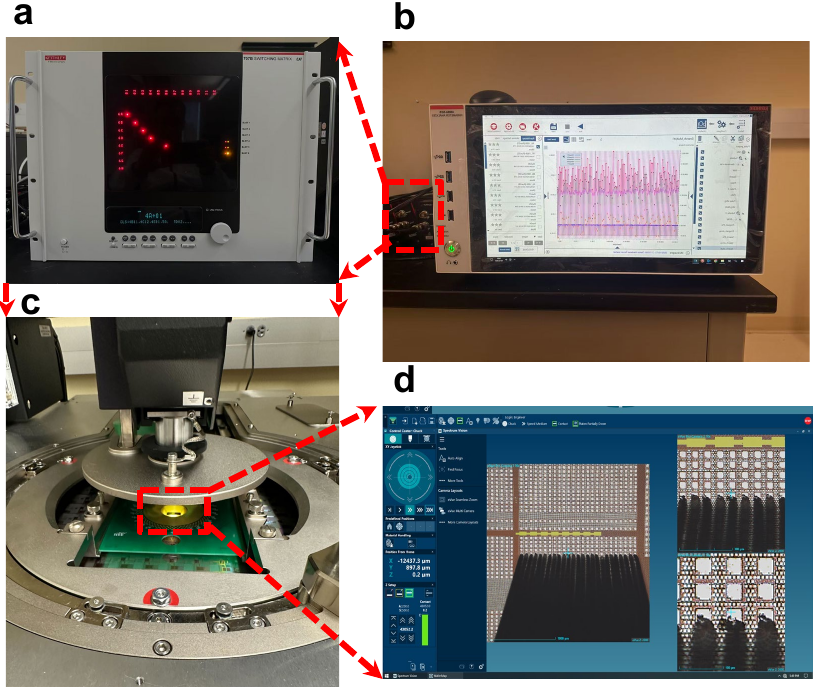}
    \caption{\textbf{FDSOI-FeFET array measurement platform.} \textbf{a,} Switching matrix used for device selection during ACAM programming. \textbf{b,} Keithley 4200 for pulse generation during ACAM array write and read operations. \textbf{c,} Probe card for establishing electrical connection to the wafer. \textbf{d,} Optical microscope image showing the ACAM array on the wafer.}
    \label{fig:setup}
\end{figurehere}

\newpage
\begin{figurehere}
   \centering
    \includegraphics[scale=1,width=1\textwidth]{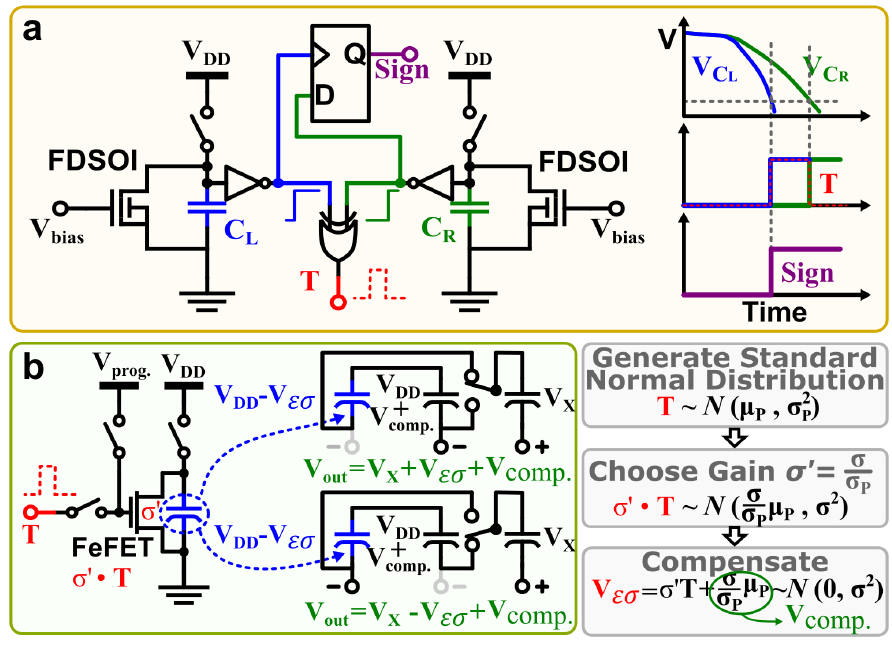}
    \caption{\textbf{Gaussian voltage generation circuit.} \textbf{a,} The Gaussian-distributed read currents from two different FDSOI-FeFET devices are used to charge the capacitors in the left and right branches respectively. The capacitors are then discharged at different rates and fed into an XOR gate, generating output pulses with Gaussian-distributed widths. \textbf{b,} The $\sigma$-scaling is implemented by applying the Gaussian pulse to the gate of a FeFET, where the scaling factor is stored in its threshold voltage. The FeFET modulates the charging of a capacitor accordingly. Subsequently, three capacitors carrying Gaussian voltage $V_{\mathrm{DD}} - V_{\epsilon\sigma}$, compensation voltage $V_{\mathrm{DD}} - V_{\mathrm{COMP}}$, and query voltage $V_{X}$ are connected together to generate the stochastic input $V_{X}$ + $V_{\epsilon\sigma}$ for Gaussian inference. Equation of each steps are shown in the figure.}
    \label{fig:GRNG_circuit}
\end{figurehere}

\newpage
\begin{figurehere}
   \centering
    \includegraphics[scale=1,width=0.7\textwidth]{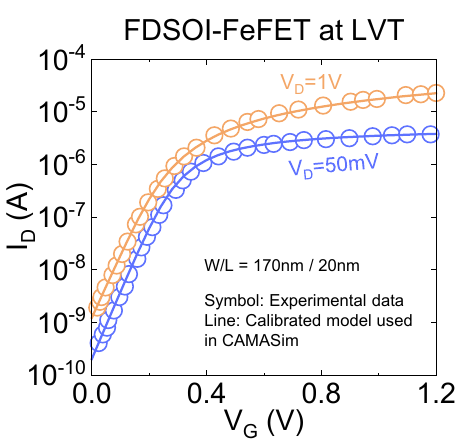}
    \caption{\textbf{FDSOI-FeFET model used in CAMASim} The experimental data for the FDSOI-FeFET were obtained at $V_{DS}=1~\mathrm{V}$ and $V_{DS}=50~\mathrm{mV}$ from a scaled device with $W/L = 170~\mathrm{nm}/20~\mathrm{nm}$, and were subsequently calibrated for ACAM functional simulation in CAMASim..}
    \label{fig:model}
\end{figurehere}

\begin{figurehere}
   \centering
    \includegraphics[scale=1,width=0.95\textwidth]{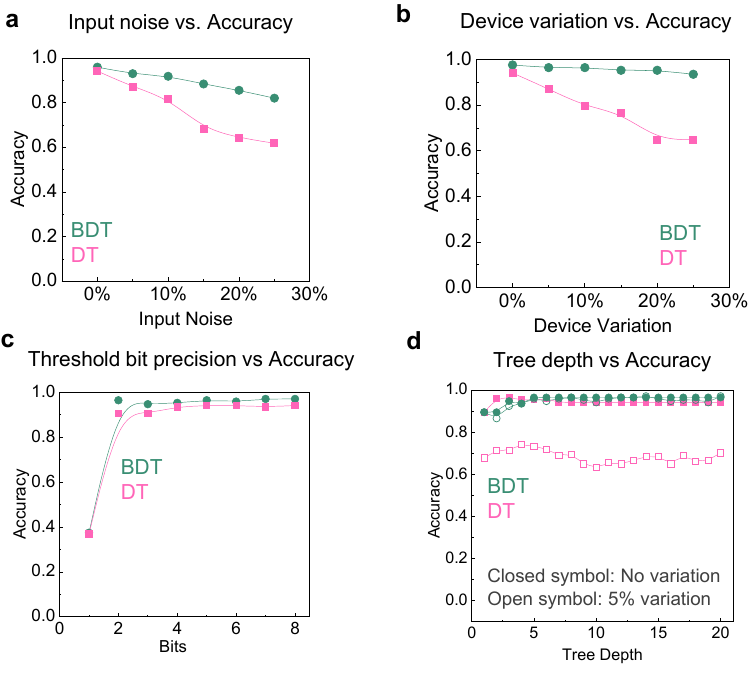}
    \caption{\textbf{Variation robustness of BDT in breast cancer dataset.} \textbf{a,} Input noise resilience evaluated on the Breast Cancer dataset. \textbf{b,} Device-level variation analysis using CAMASim. \textbf{c,} Simulation results showing classification accuracy as a function of threshold quantization bits, indicating that 2-bit precision is sufficient for breast cancer classification. \textbf{d,} Noise resilience under different tree depths.}
    \label{fig:breastcancer}
\end{figurehere}

\begin{table}[t]
\centering
\small
\begin{tabular}{l l l c c}
\hline
Model & Accelerator & Process & Latency (ns\,dec$^{-1}$) & Energy (nJ\,dec$^{-1}$) \\
\hline

DT  & Intel Core i9-14900 & Intel 7 (10 nm-class)\cite{intel7_wiki} 
& $1.02 \times 10^{3}$ & $1.08 \times 10^{5}$ \\

BDT & Intel Core i9-14900 & Intel 7 (10 nm-class)\cite{intel7_wiki} 
& $3.62 \times 10^{6}$ & $6.48 \times 10^{7}$ \\

BDT & NVIDIA RTX 4060 & TSMC 4N (5 nm-class)\cite{ada_lovelace} 
& $5.96 \times 10^{4}$ & $5.24 \times 10^{5}$ \\

DT  & ACAM (this work) & 28 nm 
& $8.21$ & $7.24$ \\

BDT & ACAM (this work) & 28 nm 
& $1.24 \times 10^{3}$ & $9.21 \times 10^{2}$ \\

\hline
\end{tabular}

\caption{\textbf{Latency and energy comparison.} 
The CPU and GPU baselines are evaluated using an Intel Core i9-14900 processor and an NVIDIA RTX 4060 GPU\cite{intel_i9_2024,nvidia_4060_2023}, fabricated in Intel 7 (10 nm-class) and TSMC 5 nm-class (4N) technologies, respectively\cite{intel7_wiki,ada_lovelace}. 
The proposed FDSOI-FeFET ACAM design is simulated using the GlobalFoundries 28 nm PDK in Cadence Virtuoso. 
The reported ACAM results include the ACAM array, GRNG, and peripheral circuits.}

\label{tab:latency_energy}
\end{table}

\clearpage
\putbib
\end{bibunit}

\end{document}